\documentclass[12pt]{article}
\pdfoutput=1

\usepackage{amsmath,amssymb,amscd}
\usepackage{listings}
\usepackage{caption}
\usepackage{dsfont}
\usepackage{slashed}
\usepackage{color}

\usepackage[pdftex]{graphicx}
\usepackage{epstopdf}
\usepackage{subfigure}
\usepackage{epsfig}
\usepackage{listings}
\usepackage{caption}
\usepackage{cite}

\usepackage{multirow}

\newcommand{\bea}{\begin{align}}
\newcommand{\eea}{\end{align}}
\newcommand{\beq}{\begin{equation}}
\newcommand{\eeq}{\end{equation}}
\newcommand{\nbea}{\begin{align*}}
\newcommand{\neea}{\end{align*}}
\newcommand{\nbeq}{\begin{equation*}}
\newcommand{\neeq}{\end{equation*}}
\newcommand{\bear}{\begin{eqnarray}}  
\newcommand{\eear}{\end{eqnarray}}  

\numberwithin{equation}{section}

\begin{document}

\def\thefootnote{\fnsymbol{footnote}}

\begin{flushright}
{\tt KCL-PH-TH/2015-20, LCTS/2015-10, CERN-PH-TH/2015-098} \\
{\tt ACT-04-15, MIFPA-15-11}
\end{flushright}

\vspace{0.7cm}
\begin{center}
{\bf {\Large An Updated Historical Profile of the Higgs Boson}}
\end{center}

\medskip

\begin{center}{\large
{\bf John~Ellis}$^{a}$,
{\bf Mary~K.~Gaillard}$^{b}$ and
{\bf Dimitri~V.~Nanopoulos}$^{c}$
}
\end{center}

\begin{center}
{\em $^a$Theoretical Particle Physics and Cosmology Group, Department of
  Physics, King's~College~London, London WC2R 2LS, United Kingdom;\\
Theory Division, CERN, CH-1211 Geneva 23,
  Switzerland}\\[0.2cm]

{\em $^b$Department of Physics, University of California and
Theoretical Physics Group, \\
Bldg. 50A5104, Lawrence Berkeley National Laboratory
Berkeley, CA 94720, USA}\\[0.2cm]

{\em $^c$George P. and Cynthia W. Mitchell Institute for Fundamental Physics and Astronomy,
Texas A\&M University, College Station, TX 77843, USA;\\
Astroparticle Physics Group, Houston Advanced Research Center (HARC), Mitchell Campus, Woodlands, TX 77381, USA;\\
Academy of Athens, Division of Natural Sciences,
28 Panepistimiou Avenue, Athens 10679, Greece}\\[0.2cm]\end{center}

\vspace{0.7cm}

\centerline{\bf ABSTRACT}
~~\\
\noindent  
The Higgs boson was postulated in 1964, and phenomenological studies
of its possible production and decays started in the early 1970s,
followed by studies of its possible production in $e^+ e^-$, ${\bar p}
p$ and $pp$ collisions, in particular. Until recently, the most
sensitive searches for the Higgs boson were at LEP between 1989 and
2000, which were complemented by searches at the Fermilab
Tevatron. Then the LHC experiments ATLAS and CMS entered the hunt,
announcing on July 4, 2012 the discovery of a ``Higgs-like'' particle
with a mass of about 125~GeV. This identification has been supported
by subsequent measurements of its spin, parity and coupling properties.  It was
widely anticipated that the Higgs boson would be accompanied by
supersymmetry, although other options, like compositeness, were not
completely excluded. So far there are no signs any new physics, and
the measured properties of the Higgs boson are consistent with the
predictions of the minimal Standard Model. This article reviews some
of the key historical developments in Higgs physics over the past
half-century.

\vspace{0.7cm}
\begin{flushleft}
April 2015
\end{flushleft}

\newpage

\section{Introduction}

The Standard Model of particle physics codifies the properties and
interactions of the fundamental constituents of all the visible matter
in the Universe. It describes successfully the results of myriads of
accelerator experiments, some of them to a very high degree of
precision.  However, for quite some time the Standard Model resembled
a jigsaw puzzle with one piece missing: the Higgs boson. It, or
something capable of replacing it, was essential for the calculability
of the Standard Model and its consistency with experimental data.  The
last piece of the puzzle, at times (somewhat dubiously) termed the
``Holy Grail'' of particle physics, or even the ``God Particle'', was
finally put into place with the July 4, 2012, announcement of the
discovery~\cite{July4} at the CERN Large Hadron Collider (LHC) of a
``Higgs-like'' particle at a mass of approximately 125 GeV.
Subsequently, measurements of its properties by the ATLAS and CMS
collaborations have shown more detailed consistency with
predictions for the Higgs boson of the Standard Model, but searches for
possible discrepancies indicative of new physics beyond the Standard Model
are continuing.

The existence of the Higgs boson was first postulated in
1964~\cite{H2}, following earlier theoretical work that introduced
spontaneous symmetry breaking into condensed-matter~\cite{Anderson}
and particle physics~\cite{Nambu,EB,H1}. It was incorporated into the
Standard Model in 1967~\cite{Weinberg,Salam}, and shown in
1971~\cite{tHV} to lead to a calculable and predictive unified theory
of the weak and electromagnetic interactions. With
the discovery of neutral currents in 1973~\cite{NC}, the discovery
of charmonium in 1974~\cite{Jpsi}, the discoveries of the $W^\pm$ and
$Z^0$ particles in 1983~\cite{WZ} and subsequent detailed measurements, the predictions of the Standard Model have
been crowned with a series of successes.

Already in 1975, before the experimental discovery of charm was
confirmed, the authors considered that the discovery of the Higgs boson would
be the culmination of the experimental verification of the Standard
Model, and we published a paper outlining its phenomenological
profile~\cite{EGN}. At the time, the ideas of spontaneously-broken
gauge theories were still generally regarded as quite hypothetical,
and the Higgs boson was not on the
experimental agenda. However, its star rose over the subsequent years,
first in $e^+ e^-$ collisions~\cite{EG76} and subsequently in ${\bar
  p} p$ and $pp$ collisions~\cite{GGMN,GNY}, until it became widely
(though incompletely) perceived as the primary objective of
experiments at the LHC.  The 2012 ATLAS and CMS discovery has finally
provided closure on half a century of theoretical conjecture,
and set the stage for a new phase of searches for physics
beyond the Standard Model.

In this paper we trace the trajectory of the Higgs boson from its
humble theoretical origins, through its rise to phenomenological
prominence, to its experimental apotheosis. However, its discovery
raises as many questions as it answers.

\section{Prehistory}

The physicist's concept of the vacuum does not correspond to the naive idea of
`empty' space. Instead, a physicist recognizes that even in the absence of physical 
particles there are quantum effects due to `virtual' particles
fluctuating in the vacuum. For a physicist, the vacuum
is the lowest-energy state, after taking these quantum effects into account. This
lowest-energy state may not possess all the symmetries of the underlying equations
of the physical system, a phenomenon known as `spontaneous' symmetry breaking,
or `hidden' symmetry.

This mechanism of spontaneous symmetry breaking first came to prominence in
the phenomenon of superconductivity, as described in the theory of Bardeen, Cooper and Schrieffer~\cite{BCS}.
According to this theory, the photon acquires an effective mass when it propagates through certain materials
at sufficiently low temperatures, as discussed earlier by Ginzburg and Landau~\cite{GL}. In free space, 
the masslessness of the photon is guaranteed by Lorentz invariance and U(1) gauge symmetry. 
A superconductor has a well-defined rest frame, so Lorentz invariance is broken explicitly. However, the gauge
symmetry is still present, though `hidden' by the condensation of Cooper pairs of electrons~\cite{Cooper} in the lowest-energy
state (vacuum). It was explicitly shown by Anderson~\cite{Anderson} how the interactions 
with the photon of the Cooper pairs inside a superconductor caused the former to acquire an effective mass.

The idea of spontaneous symmetry breaking was introduced into particle physics by Nambu~\cite{Nambu} in 1960. 
He suggested that the small mass and low-energy interactions of pions could be understood as a reflection of a
spontaneously-broken global chiral symmetry, which would have been exact if the up and down quarks were massless. His
suggestion was that light quarks condense in the vacuum, much like the Cooper pairs of superconductivity.
When this happens, the `hidden' chiral symmetry causes the pions' masses to vanish, and fixes their low-energy couplings 
to protons, neutrons and each other.

\begin{figure}[h!]
\centering
\includegraphics[scale=0.4]{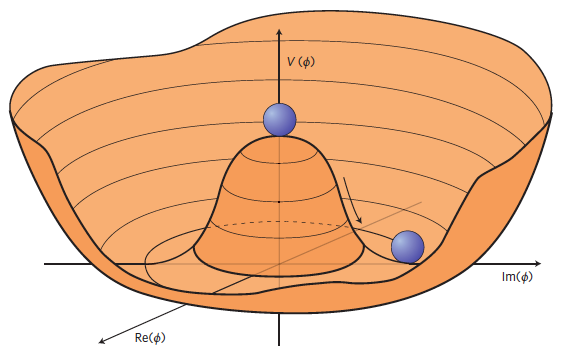}
\caption{\it A prototypical effective `Mexican hat' potential that leads to `spontaneous' symmetry breaking.
The vacuum, i.e., the lowest-energy state, is described by a randomly-chosen point around the bottom of the
brim of the hat. In a `global' symmetry, movements around the bottom of the hat corresponds to a massless spin-zero
`Nambu-Goldstone' boson~\cite{Nambu,Goldstone}. 
In the case of a local (gauge) symmetry, as was pointed out by 
Englert and Brout~\cite{EB}, by Higgs~\cite{H2}
and by Guralnik, Hagen and Kibble~\cite{GHK}, 
this boson combines with a massless spin-one boson to yield a
massive spin-one particle. The Higgs boson~\cite{H2} is a 
massive spin-zero particle corresponding to quantum
fluctuations in the radial direction, oscillating between the centre and the side of the hat.
}
\label{fig:hat}
\end{figure}

A simple model of spontaneous global U(1) symmetry breaking was introduced by Goldstone~\cite{Goldstone} in 1961,
with a single complex field $\phi$ as illustrated in
Fig.~\ref{fig:hat}.  The effective potential
\beq V(|\phi|) =
\frac{\lambda}{4} \, (|\phi|^2 - v^2)^2\label{Vphi}\eeq 
\noindent
is unstable at the origin where $\langle |\phi| \rangle = 0$. Instead, the
lowest-energy state, the vacuum, is at the bottom of the brim of the
`Mexican hat', with 
\beq\langle |\phi| \rangle = v\ne 0 \, .\label{phivev}\eeq 
\noindent
The phase of $\phi$ is, however, not determined, and all choices are
equivalent with the same energy. The system must choose some
particular value of the phase, but changing the phase would cost no
energy. Hence the system has a massless degree of freedom
corresponding to rotational fluctuations of the field around the brim
of the Mexican hat. It is a general theorem, proven later in 1961 by
Goldstone, Salam and Weinberg~\cite{GSW} that spontaneous breaking of
a global symmetry such as chiral symmetry must be accompanied by the
appearance of one or more such Nambu-Goldstone bosons.

However, this
is not necessarily the case if it is a gauge symmetry that is broken,
as in the non-relativistic case of superconductivity~\cite{Anderson}.
Anderson conjectured that it should be possible to extend this
mechanism to the relativistic case, as did Klein and Lee~\cite{KL},
but it was argued by Gilbert~\cite{Gilbert} that Lorentz invariance
would forbid this.

\section{And then there was Higgs}

Spontaneous breaking of gauge symmetry was introduced into particle
physics in 1964 by Englert and Brout~\cite{EB}, followed independently
by Higgs~\cite{H1,H2}, and subsequently by Guralnik, Hagen and
Kibble~\cite{GHK}. They demonstrated how one could dispose
simultaneously of two unwanted massless bosons, a spinless Nambu-Goldstone
boson and a gauge boson of an exact local symmetry, by combining them
into a single massive vector boson in a fully relativistic theory.
The two polarization states of a massless vector boson are combined
with the single degree of freedom of a spin-zero particle to yield the
three degrees of freedom of a massive spin-one particle $V$ with mass:
\beq m_V = g_V \, \frac{v}{\sqrt{2}} \, ,\label{Vmass}\eeq
\noindent
where $g_V$ is the corresponding gauge coupling constant.

Englert and Brout~\cite{EB} considered explicitly a non-Abelian Yang-Mills theory, assumed the formation of a vacuum
expectation value (vev) of a non-singlet scalar field, and used a diagrammatic approach to demonstrate mass
generation for the gauge field. The first paper by Higgs~\cite{H1} demonstrated that gauge symmetry
provides a loophole in the `no-go' theorem of Gilbert mentioned above, and his second paper~\cite{H2} exploited this loophole
to demonstrate mass generation in the Abelian case. The subsequent paper 
by Guralnik, Hagen and Kibble~\cite{GHK} referred in its text to the Englert/Brout and Higgs papers, and
also demonstrated mass generation in the Abelian case.

The second paper by Higgs~\cite{H2} is the only one of the 1964 papers
to mention explicitly [his equation (2b)] the existence of a massive
scalar particle associated with the curvature of the effective
potential (\ref{Vphi}) that determines the vev $v$ of the charged field:
\beq m_H = \sqrt{2\lambda} \, v \, .\label{Hmass}\eeq 
\noindent
Englert and Brout~\cite{EB} do not discuss the spectrum of physical
scalars, whilst Guralnik, Hagen and Kibble~\cite{GHK} mention a
massless scalar that decoupled from the massive excitations in their
model.

Also worthy of note is a remarkable paper written in ignorance of these papers by Migdal and Polyakov in 1965~\cite{MP},
while they were still students~\footnote{It was finally published in 1966
after a substantial delay caused by the scepticism of Soviet academicians.}, in which they
discuss partial spontaneous symmetry breaking in the non-Abelian case.
The year 1966 also saw a further important paper by Higgs~\cite{H3},
in which he discussed in detail the formulation of the spontaneously-broken Abelian theory.
In particular, he derived explicitly the Feynman rules for processes involving what has come to be
known as the massive Higgs boson, discussing its decay into 2 massive vector
bosons, as well as vector-scalar and scalar-scalar scattering. Another important paper by Kibble~\cite{Kibble}
discussed in detail the non-Abelian case, including partial spontaneous symmetry
breaking, and also mentioned the appearance of massive scalar bosons
\`a la Higgs.

The next important step was the incorporation by
Weinberg~\cite{Weinberg} and by Salam~\cite{Salam} of non-Abelian
spontaneous symmetry breaking into Glashow's~\cite{Glashow} unified
SU(2) $\times$ U(1) model of the weak and electromagnetic
interactions. The paper by Weinberg was the first to observe that the
scalar field vev could also give masses to fundamental fermions $f$ that
are proportional to their coupling to the Higgs boson:
\beq m_f = g_{f\bar f H} \, v \, .\label{fmass}\eeq
However, the seminal papers on spontaneous breaking of gauge symmetries and
electroweak unification were largely ignored by the particle physics
community until the renormalizability of spontaneously-broken gauge theories was demonstrated by
't Hooft and Veltman~\cite{tHV}. These ideas then joined the mainstream very rapidly, thanks
in particular to a series of influential papers by B.~W.~Lee and collaborators~\cite{LZJ,FLS}.

\section{A Phenomenological Profile of the Higgs Boson}

B.~W.~Lee also carries much of the responsibility for calling the
Higgs boson the Higgs boson, mentioning repeatedly `Higgs scalar
fields' in a review talk at the International Conference on
High-Energy Physics in 1972~\cite{Lee}. However, in the early 1970s
there were only a few suggestions how to constrain or exclude the
possible existence of a physical Higgs boson. One paper considered the
possible effect of Higgs exchange on neutron- and deuteron-electron
scattering and derived a lower bound $m_H > 0.6$~MeV~\cite{nue}, and
another constrained Higgs emission from neutron stars, yielding the
lower bound $m_H > 0.7$~MeV~\cite{SS}.  There was also a theoretical
discussion of possible Higgs production in $0^+ \to 0^+$ nuclear
transitions~\cite{RSW}, and its non-observation in excited $^{16}$O
and $^4$He decays led to the Higgs being excluded in the mass range
$1.03~{\rm MeV} < m_H < 18.3$~MeV~\cite{KWB}. In parallel, data on
neutron-nucleus scattering were used to constrain $m_H >
15$~MeV~\cite{BE}.  The two latter were the strongest limits obtained
in this period.

This was the context in which we embarked in 1975 on the first systematic study of possible Higgs phenomenology~\cite{EGN}.
Neutral currents had been discovered~\cite{NC}, and the J/$\psi$ particle~\cite{Jpsi} was thought to be charmonium, though
doubts remained and the discovery of open charm still lay in the future. The search for the intermediate vector
bosons $W^\pm$ and $Z^0$ was appearing on the experimental agenda, but the CERN ${\bar p}p$
collider that was to discover them had not yet been proposed. However, it seemed to us that the clinching
test of the spontaneous symmetry-breaking paradigm underlying the Standard Model would be discovering
the Higgs boson.

To this end, we considered the decay modes of the Higgs boson if it weighed up to 100~GeV, calculating for
the first time the loop-induced Higgs decays to photon pairs~\cite{EGN}. The dominant mechanism for this decay is an
anomalous triangle diagram with a $W^\pm$ loop, and there are also subdominant diagrams with
massive quarks~\footnote{We did not calculate these: the $t$ and $b$ had not been discovered at that time.}~\cite{Djouadi}:
\beq
\Gamma(H \to \gamma \gamma) \; = \; \frac{G_\mu \alpha^2 m_H^3}{128 \sqrt{2} \pi^3}
\left| A_1 (r_W) + \sum_f N_c Q_f^2 A_{1/2} (r_f)\right|^2 \, ,
\label{Djouadi}
\eeq
where
\beq
A_1(r) \; \equiv \; -\left[2 r^2 + 3 r + 3(2 r - 1)f(r)\right]/r^2, \; \; A_{1/2}(r) \; \equiv\;  2 \left[r + (r - 1) f(r) \right]/r^2
\eeq
and $f (r) \equiv \arcsin^2 \sqrt{r}$ for $r_{W, f} \equiv m_H^2/4m_{W, f}^2$. We also estimated the cross
sections for many different mechanisms for producing the Higgs boson, intending to cover the full allowed mass range
from ${\cal O}(15)$~MeV upwards. In addition to considering the production of a relatively light Higgs boson in
hadron decays and interactions, we also considered production in $e^+ e^-$ collisions, including $Z^0$ decays and Higgs-strahlung 
processes such as $e^+ e^- \to Z^0 + H$~\cite{EGN}~\footnote{The latter processes were also considered
independently in~\cite{Bj} and~\cite{IK}, respectively.}.

Back in 1975, the likelihood of a definitive search for the Higgs boson seemed somewhat remote. That was
why, rather tongue-in-cheek, we closed our paper~\cite{EGN} with the following modest words: {\it ``We should
perhaps finish our paper with an apology and a caution. We apologize to experimentalists for having no idea what is the
mass of the Higgs boson, ..., and for not being sure of its couplings to other particles, except that they are
probably all very small. For these reasons, we do not want to encourage big experimental searches for the Higgs boson,
but we do feel that people doing experiments vulnerable to the Higgs boson should know how it may turn up."}

In those early days, there was very little theoretical guidance as to
the possible mass of the Higgs boson, which is one reason why these
early studies also included very low Higgs masses. One possibility
that attracted attention was that the Higgs mass was entirely due to
quantum corrections, which would have yielded $m_H \sim 10$~GeV in the
absence of heavy fermions~\cite{CW}~\footnote{This was long before it
  was recognized that the top quark weighed $> m_W$.}. At the other
end of the mass scale, it was emphasized that the Higgs
self-interactions would become strong for $m_H \sim
1$~TeV~\cite{heavyH}.

\section{Searches for the Higgs Boson at LEP}

In addition to~\cite{EGN}, there was an early discussion of searches
for the Higgs boson in $e^+ e^-$ collisions in~\cite{EG76}.  There are
three important processes for producing the Higgs boson at an $e^+
e^-$ collider: in $Z^0$ decay - $Z^0 \to H + {\bar
  f}f$~\cite{EG76,Bj}, in association with the $Z^0$ - $e^+ e^- \to
Z^0 + H$~\cite{EGN} with the cross-section~\cite{IK,LQT}
\beq
\sigma(e^+ e^- \to Z + H) \; = \; \frac{\pi \alpha^2}{24} \left( \frac{2 p}{\sqrt{s}}\right) \left( \frac{p^2 + 3 m_Z^2}{(s - m_Z^2)^2} \right)
\left( \frac{1 - 4 \sin^2\theta_W + 8 \sin^4 \theta_W}{\sin^2 \theta_W (1 - \sin^2 \theta_W)^2} \right) \, ,
\eeq
where $p$ is the momentum of the final-state particles, and via $W^+ W^-$ or $Z^0 Z^0$ fusion -
$e^+ e^- \to {\bar \nu} H \nu, e^+ H e^-$~\cite{JP}.  The direct
process $e^+ e^- \to H$ is negligible because--see (\ref{fmass})--of the
small $H$ coupling to $e^+ e^-$~\cite{EGN}, though the corresponding
reaction at a muon collider, $\mu^+ \mu^- \to H$, may be
interesting~\cite{mumu}.  We also note that high-intensity lasers may
be able to convert an $e^+ e^-$ collider into a high-luminosity
$\gamma \gamma$ collider, which might also be an interesting Higgs
factory~\cite{gammagamma}.

It should be noted that in the early CERN reports on LEP physics the only
discussions of Higgs production were theoretical~\cite{EG76,LesHouches}. The concerns of our
experimental colleagues lay elsewhere, and these early CERN reports contain no experimental
discussions of possible searches for the Higgs boson. The first written experimental discussion of
which we are aware was in an unpublished 1979 report for ECFA~\cite{ECFA}, compiled by a joint working group
of theorists and experimentalists. This was followed in 1985 by a more detailed study by a
joint theoretical and experimental working group in a CERN report~\cite{LEP85} published in 1986.
Thereafter, Higgs searches were firmly in the experimental sights of the LEP collaborations, as seen
in later CERN reports on LEP physics~\cite{sights}.

\begin{figure*}[htb]
\begin{center}
\resizebox{7cm}{!}{\includegraphics{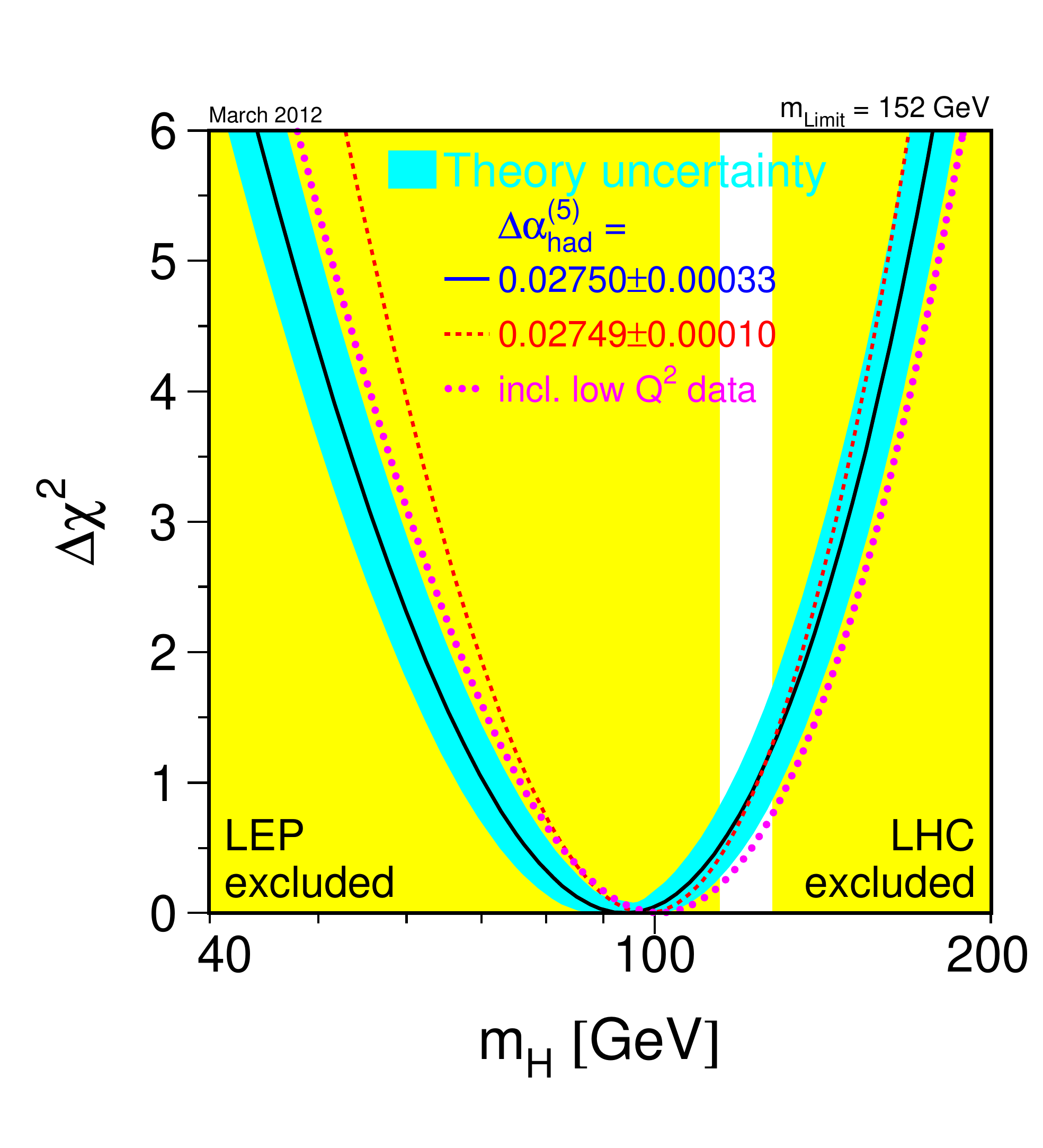}}
\end{center}
\caption{\it A compilation of information about the possible mass of
  the Higgs boson just prior to the LHC discovery~\protect\cite{LEPEWWG}. The yellow-shaded regions
  had been excluded by searches at LEP~\protect\cite{LEPH}, the
  Tevatron collider~\protect\cite{TeVH} and the
  LHC~\protect\cite{ATLASH,CMSH}. The black line (and the blue band)
  is the $\chi^2$ function for the precision electroweak data as a
  function of $m_H$ (and its theoretical uncertainty). The dotted
  lines are obtained using alternative treatments of the precision
  electroweak data~\protect\cite{LEPEWWG}.  }
\label{fig:blueband}
\end{figure*}

In parallel with the searches for the Higgs boson, notably that for
the process $Z^0 \to H + {\bar f}f$~\cite{EG76,Bj} at LEP~1 and that for
$e^+ e^- \to Z^0 + H$~\cite{EGN,IK,LQT} at LEP~2, the high-precision
electroweak data obtained at LEP, the SLC and elsewhere made it
possible for the first time to estimate the possible mass of the Higgs boson
within the framework of the Standard Model~\cite{EF}. The dominant $m_H$-dependent
corrections had been calculated earlier~\cite{Veltman}, but the possibility of using
them in conjunction with LEP data to constrain $m_H$ was not discussed
before LEP start-up, perhaps because the precision of LEP data exceeded
all previous expectations. The constraint on $m_H$ was relatively weak before
the top quark was discovered~\cite{Tevatront} (with a mass that agreed with predictions
based on electroweak data), but the measurement of $m_t$ allowed more
accurate estimates of $m_H$ to be made. Values $< 300$~GeV were favoured
already in the early days of such studies~\cite{EF}, which have now matured to indicate
that $m_H \sim 100 \pm 30$~GeV~\cite{LEPEWWG}. This constraint is combined with the (negative)
results of the direct Higgs searches in Fig.~\ref{fig:blueband}.

The non-appearance of the Higgs boson in searches in $Z^0$ decays
required $m_H > 58$~GeV~\cite{LEP1}.  Thereafter, successive increases
in the LEP energy during the LEP~2 era prompted recurrent hopes that
the Higgs discovery might lie just around the corner, but instead the
lower limit on $m_H$ kept rising inexorably. Finally, in 2000 the LEP
centre-of-mass energy was pushed to 206~GeV, and a few Higgs-like
events were observed, corresponding to a mass $\sim
115$~GeV~\cite{115}. To the disappointment of many physicists, it was
not possible to push the LEP energy higher, and the difficult decision
was taken to shut LEP down at the end of the year 2000, leaving the
lower limit $m_H > 114.4$~GeV at the 95\%~CL~\cite{LEPH}. There was
much speculation at the time that this decision forced LEP to miss out
on the Higgs discovery. However, with $m_H \approx 125$~GeV as has now
been established~\cite{July4} at the LHC, substantial extra
investment in accelerating cavities would have been necessary back in
the 1990s in order to be able to push LEP to sufficiently high
energies to produce it.

\section{Searches for the Higgs Boson at Hadron Colliders}

The production of the Higgs boson at hadron colliders is more problematic than
in $e^+ e^-$ collisions. On the one hand, the backgrounds from other physical processes are large,
and on the other hand direct production via the dominant quark constituents in
the proton is small, because they have very small masses~\cite{EGN}. There is in addition, 
however, production by gluon-gluon fusion via anomalous triangle diagrams~\cite{RU}:
as first discussed in~\cite{GGMN}:
\beq
\frac{d \sigma}{d y} \; =  \; \left(\frac{\alpha_s}{\pi}\right)^2 \left(\frac{\pi G_\mu}{288 \sqrt{2}}\right) r G(\sqrt{r} e^y) G(\sqrt{r} e^{-y}) \, ,
\label{GGMN}
\eeq
where $y$ is the rapidity, $r \equiv m_H^2/s$ and $G$ is the gluon distribution function within the proton.
This is the dominant Higgs production mechanism at the LHC, and it is ironic that this and
one of the most distinctive Higgs decays, that into $\gamma \gamma$~\cite{EGN}, are both due to
similar quantum effects.
Another important production mechanism is Higgs-strahlung in association with a $W^\pm$ or $Z^0$, 
as was first discussed in~\cite{GNY}, which was the dominant Higgs production mechanism at the Tevatron.
A third important mechanism is $W^+ W^-$ (and $Z^0 Z^0$)
fusion, as first discussed in~\cite{CD}.

A comprehensive theoretical survey of the new physics possibilities at the Superconducting SuperCollider (SSC)
was provided in~\cite{EHLQ}, and the search for the Higgs boson naturally took pride of place. The same
was true in the survey of the new physics possibilities at the LHC provided in~\cite{EGK}, which also
included Higgs production in association with a ${\bar t} t$ pair~\cite{Zoltan}. In anticipation of Higgs 
searches at the SSC, in particular, a comprehensive survey of the theory and phenomenology of the
Higgs boson was published~\cite{HHG}. It served as the Bible for many subsequent Higgs hunters,
also at LEP and the LHC following the much-lamented demise of the SSC~\cite{Djouadi}.

After the shutdown of LEP, the lead in Higgs searches was taken by the
CDF and D0 experiments, at the Tevatron collider, where the dominant
production mechanism was Higgs-strahlung in association with a $W^\pm$
or $Z^0$~\cite{GNY}. As the analyzed Tevatron luminosity accummulated,
CDF and D0 became able to exclude a range of Higgs masses between 156
and 177~GeV~\cite{TeVH}, as well as a range of lower masses in the
range excluded by LEP. There was a small excess of Higgs candidate
events in a range around 130 to 140~GeV, though not strong enough to
be considered a hint, let alone significant evidence. Since the most
important Higgs decay channels for the Tevatron experiments are $H \to
{\bar b}b$ and $W^+ W^-$, which have relatively poor mass resolution,
the excess did not provide much information what value $m_H$ might
have. Unfortunately, the Tevatron was shut down in September 2011,
before it could realize its full potential for Higgs searches.

The LHC started producing collisions in late 2009, initially at low
energies, and starting at 7~TeV in the centre of mass in March 2010. By
the end of 2011, the ATLAS and CMS experiments had each accumulated
$\sim 5$/fb of data, and in 2012 they accumulated $\sim 20$/fb of data at
8~TeV in the centre of mass. Already at the end of 2011 optimists could see a hint
in their data of a new particle with a mass $\sim 120$ to $125$~GeV, and this was
followed by the announcement on July 4th, 2012 of the discovery of a new particle
weighing $\sim 125$~GeV that resembled, {\it prima facie}, a Higgs boson.
Later analyses of the Tevatron data subsequently supported the
LHC discovery, albeit with much lower level of significance~\cite{TevatronH}.

\section{Is it really a/the Higgs boson?}

Although this new particle was widely expected to be a/the Higgs boson, and its measured
mass~\cite{MHjoint}
\beq
m_H \; = \; 125.09 \pm 0.24 \; {\rm GeV}
\label{measuredmH}
\eeq
is certainly consistent with the previous indications from precision electroweak and other
 data, it was important
to check its properties and exclude possible alternatives. The following is a check-list
of some properties that needed to be verified:

$\bullet$ What is its spin? A Higgs boson must have spin 0, and this is consistent with
its observation in the $\gamma \gamma$ final state, which excludes spin 1. However,
integer spins $\ge 2$ remained a possibility, albeit unexpected.

$\bullet$ Is it scalar or pseudoscalar? In the Standard Model the Higgs would necessarily
be a scalar, but many models have a family of Higgs-like particles, with at least one being
a pseudoscalar, e.g., supersymmetry, and these could mix in the presence of CP violation.

$\bullet$ Is it elementary to the same extent as the other Standard Model particles, or
does it show signs of being a composite particle, like a Cooper pair or a pion?

$\bullet$ Does it couple to other particles in proportion to their masses? This would be 
a `smoking gun' for the new particle's connection with the origin of particle masses.

$\bullet$ Are quantum effects in the new particle couplings consistent with calculations
within the Standard Model?

As we now discuss, so far the new particle discovered by ATLAS and CMS has passed
all these tests with flying colours.

Many probes of the putative Higgs spin have been proposed and used, including kinematic correlations
in $H \to W W^*$ and $Z Z^*$ decays, the kinematics of associated $H + W/Z$ production and the
energy dependence of the $H$ production cross-section~\cite{ATLASHspin,CMSHspin,TevatronHspin}. Here we mention just one example, the
angular distribution of the final-state photons in $gg \to H \to \gamma \gamma$~\cite{Hspin}. In the case of a spin-0
particle, this angular distribution would be isotropic, but the same would not be true for a spin-2 particle
with graviton-like couplings. It would be produced by gluon pairs with parallel spins in an equal admixture
of states with spin $\pm 2$ along the collision axis. This non-trivial initial spin state would be imprinted
on the angular distribution of the decay photons:
\begin{equation}
\frac{d\sigma}{d\Omega}\ \propto\ \frac{1}{4} + \frac{3}{2}{\rm cos}^2{\theta}
+ \frac{1}{4}{\rm cos}^4{\theta} \, ,
\label{g10}
\end{equation}
and analysis of the LHC data strongly disfavours (\ref{g10}) compared with the isotropic hypothesis.
Many other tests also support spin 0 and disfavour a wide range of alternative hypotheses.

Many of the same strategies also distinguish between the scalar and pseudoscalar
hypotheses. One example is provided by the distribution of the invariant mass $M_{HV}$ in associated $H + W/Z (\equiv V)$ production~\cite{H+V}.
In the Standard Model the final-state particles are produced in a relative S-wave, and the cross-section
grows like $\beta$ just above the threshold $M_{HV} \to m_H + m_V$. However, if the new particle were pseudoscalar,
the cross-section would grow as $\beta^3$ just above threshold. As a consequence,  the distributions in both $M_{HV}$
and the transverse momenta of the $H$ and $V$ would be much broader in the pseudoscalar case.
Data from the Tevatron are consistent with the scalar hypothesis, and many LHC measurements of other Higgs channels
also disfavour strongly the pure pseudoscalar hypothesis.
However, in the presence of CP violation there could be channel-dependent admixtures of pseudoscalar
couplings, so these tests should be continued.

\begin{figure}[htb]
\centering
\includegraphics[height=2.3in]{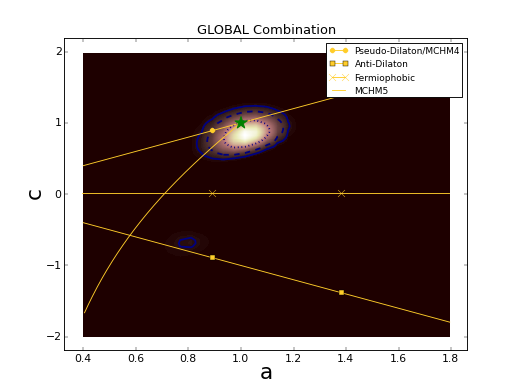}
\includegraphics[height=2.3in]{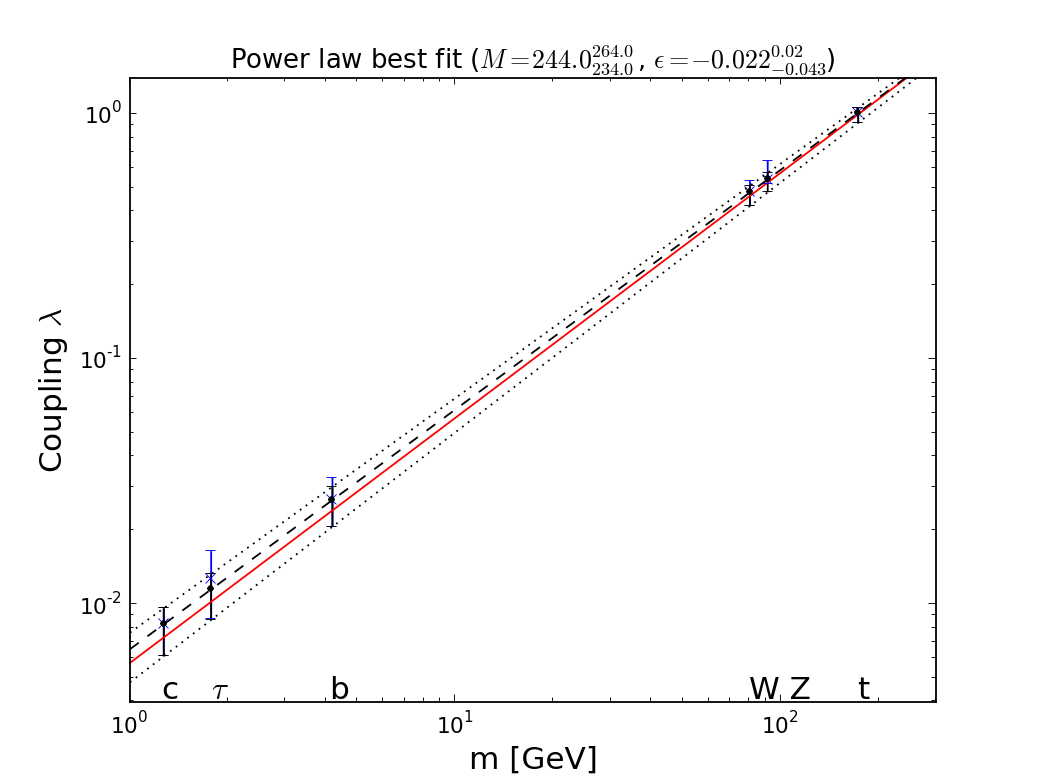}
\caption{\it Left panel: A global fit to bosonic and fermionic $H$ couplings (\protect\ref{calL}) rescaled by factors $a$ and $c$, respectively.
The SM prediction $a = c = 1$ is shown as the green star~\protect\cite{EY3}, and the yellow lines
show the possible predictions of some composite models. Right panel: A global fit to the $H$ couplings of the form (\protect\ref{Mepsilon})
(central values as dashed and
$\pm$1$\sigma$ values as dotted lines), which is very consistent
with the linear mass dependence for fermions and quadratic mass dependence for bosons (solid
red line) expected in the Standard Model~\protect\cite{EY3}.}
\label{fig:EY}
\end{figure}

One way to explore any possible composite nature of the `Higgs' boson is to parametrize its couplings to bosons and fermions as follows:
\begin{equation}
{\cal L} \; = \; \frac{v^2}{4} {\rm Tr} D_\mu \Sigma^\dagger D^\mu \Sigma \left( 1 + 2 a \frac{H}{v} + \dots \right) 
- {\bar \psi}^i_L \Sigma \psi_R \left(1 + c \frac{H}{v} + \dots \right) \, ,
\label{calL}
\end{equation}
where $\Sigma$ is a $2 \times 2$ matrix containing the 3 Goldstone fields that are `eaten' by the massive
gauge bosons appearing in the gauge-covariant derivatives $D_\mu$.
As seen in the left panel of Fig.~\ref{fig:EY}, the data are completely consistent with the Standard Model prediction $a = c = 1$:
no sign of any significant deviation that would require new strongly-interacting physics at a low energy scale,
as would be expected in a composite Higgs model.

One way to probe the mass dependence of the `Higgs' couplings is to parametrize those to fermions $\lambda_f$
and massive bosons $g_V$ in the forms~\cite{EY3}
 \begin{equation}
\lambda_f \; = \; \sqrt{2} \left(\frac{m_f}{M}\right)^{1 + \epsilon}, \; g_V \; = \; 2 \left(\frac{m_V^{2(1 + \epsilon)}}{M^{1 + 2\epsilon}}\right) \, .
\label{Mepsilon}
\end{equation}
where one would expect the power $\epsilon = 0$ and the scaling
coefficient $M = v = 246$~GeV in the Standard Model. The right panel of Fig.~\ref{fig:EY} shows the result of a fit
to the two parameters $(M, \epsilon)$ as a dashed line,
with the one-$\sigma$ range indicated by dotted lines. The points with error bars are the predictions of
the two-parameter fit, and the solid red line is the
prediction of the Standard Model. The data are completely compatible with the Standard
Model~\footnote{We note also that direct evidence has been presented for `Higgs' decays to fermions~\cite{CMSf}.}.

As already commented, it is ironic that the dominant Higgs production mechanism $gg \to H$
and one of its most prominent decay modes $H \to \gamma \gamma$ are quantum (loop) effects
due to anomalous triangle diagrams. (How else could it couple to massless particles?) The good news
is that this implies that one has good sensitivity to any possible additional massive particles
circulating in the loops. The right panel of Fig.~\ref{fig:EY} shows that these loop couplings
do not deviate significantly from the Standard Model, confirming that it works at the quantum level and
constraining possible extensions of the Standard Model.

In awarding the Nobel Prize to Francois Englert and Peter Higgs, the Royal Swedish Academy of Sciences
agreed~\cite{Nobel} with~\cite{EY3} that {\it ``beyond any reasonable doubt, it is a Higgs boson."}~\footnote{Even though this phrase was removed 
from the published version of~\cite{EY3} at the request of the referee, who considered it {\it ``unscientific"}.}. There are good
prospects that future runs of the LHC experiments will pin down further many of the Higgs couplings, 
including those to ${\bar t} t$ and $\mu^+ \mu^-$, but more detailed studies would require an $e^+ e^-$ collider -
running either at low energy in `Higgs factory' mode~\cite{ILC,TLEP},
or at higher energies where Higgs production would be more copious and more
Higgs couplings could be measured, including its self-couplings~\cite{CLIC}, or a higher-energy proton-proton collider~\cite{FCC-hh}.

\section{More Higgs, Less Higgs? More than Higgs?}

So far, we have focused on the least adventurous hypothesis of
a single Standard Model-like Higgs boson, but alternatives abound,
with most theorists expecting supplements to the minimal Higgs sector
of the Standard Model.

One of the simplest possibilities is that there are two complex doublets
of Higgs bosons, in which case there would be five physical Higgs bosons:
three neutral and two charged $H^\pm$. The most natural framework for such a possibility
is supersymmetry~\cite{susy}. In simple supersymmetric models the lightest of the the three neutral
Higgs boson often has couplings similar to those of the
Higgs boson in the Standard Model, whereas one of the heavier neutral
Higgs bosons would be a pseudoscalar.

The mass of the Higgs boson is linked to the magnitude of its self-coupling,
which would be fixed by supersymmetry in terms of the electroweak gauge
couplings. For this reason, supersymmetry predicts a restricted range for the mass
of the lightest Higgs boson:
\beq
m_H \; = \; \frac{1}{2} m_Z^2 + m_A^2 - \sqrt{(m_Z^2 + m_A^2)^2 - 4 m_Z^2 m_A^2 \cos^2 2 \beta} \, ,
\label{tree}
\eeq
at the classical (tree) level, where $m_A$ is the mass of the pseudoscalar Higgs boson $A$
and $\tan \beta$ is the ratio of the v.e.v.s of the two Higgs doublets.
Equation (\ref{tree}) gives $m_H< m_Z$. However, nearly 25 years ago it was realized that this prediction would be subject
to important radiative corrections due to the heavy top quark:
\beq
\Delta m_H^2 \; = \; \frac{3g^2}{4 \pi^2} \frac{m_t^4}{m_W^2} \ln \frac{m_{\tilde t}}{m_t} \, ,
\label{loop}
\eeq
where $g$ is the SU(2) gauge coupling and $m_{\tilde t}$ is the mass of the stop squark, which could push $m_H$ up to $\sim 130$~GeV
in simple supersymmetric models~\cite{ERZ}. In such a supersymmetric scenario there are no 
significant restrictions on the masses of the heavier Higgs bosons such as the pseudoscalar $A$.
The measured mass of the Higgs boson lies comfortably within
the range of Higgs masses favoured in simple supersymmetric models~\cite{ENOS}, and simple supersymmetric
models also predicted successfully that its couplings would be very similar to those in the Standard Model.
However, so far the LHC has found no evidence of any supersymmetric particles
and, as long as this is the case, theorists will consider other possibilities.

An alternative to an elementary Higgs boson of the type found in supersymmetric models would be that the
spontaneous breaking of the electroweak gauge symmetry is due to a condensate in
the vacuum of pairs of new, strongly-interacting fermions~\cite{TC}, analogous to the Cooper pairs
of superconductivity~\cite{Cooper}. In this case, there would in
general be a composite scalar particle that might be accessible to experiment. This would not necessarily
correspond to a strongly-interacting Higgs boson, which would
have to confront issues with the precision electroweak data. For example, if the composite scalar is a (relatively) light
pseudo-Goldstone boson of some higher-level broken symmetry, such as a larger chiral
symmetry~\cite{littleH} or approximate scale invariance, it would have weak interactions and could
mimic a Standard-Model-like Higgs boson to some extent. It would need to because, as we have seen, the data
give no indication of any deviation from the Standard Model.

One such example would be a pseudo-dilaton of approximate scale invariance~\cite{pseudoD}, which
would have tree-level couplings similar to those of the Higgs boson, but rescaled (and probably suppressed) by a
universal factor. The loop-induced couplings of the pseudo-dilaton to gluon and photon pairs might not share
this universal rescaling. This model provides a straw person to compare with the Standard
Model Higgs scenario. However, as we have seen, the data provide no encouragement for
such a scenario, and a Higgs-like
particle with suppressed couplings would not fulfill all the functions of the Higgs boson of the
Standard Model, e.g., in unitarizing $W^+ W^-$ scattering, and would have to be supplemented by
some other detectable degrees of freedom in the TeV energy range~\cite{G} none of which has been seen so far.

Another scenario for a light Higgs-like particle is the radion~\cite{radion}, the quantum of the degree of
freedom corresponding to rescaling an extra dimension. Models with extra dimensions offered
many other possibilities, including the possibility that there was no Higgs boson at all~\cite{Higgsless},
but those particular models are now out of fashion!

\begin{figure}[h!]
\centering
\includegraphics[scale=0.4]{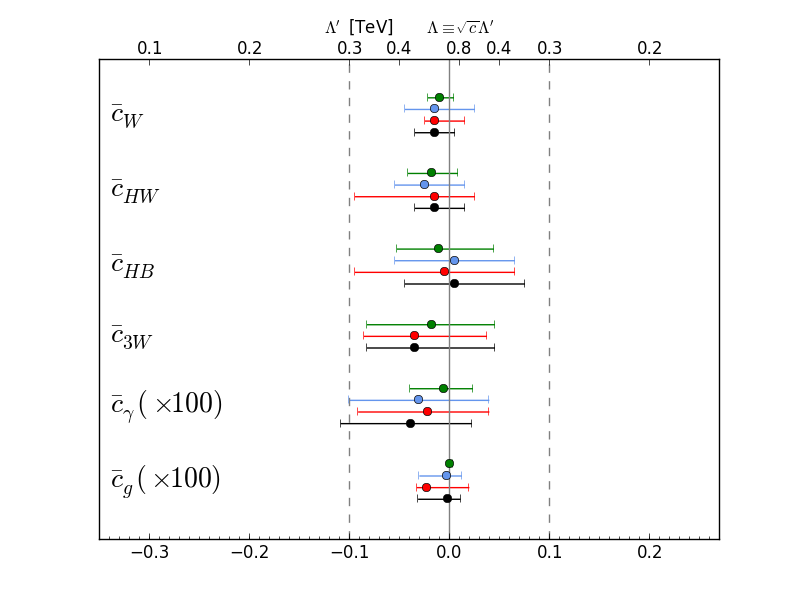}
\caption{\it The 95\% CL constraints for single-coefficient fits (green bars),
and the marginalised 95\% ranges for the
LHC Higgs signal-strength data combined with the kinematic distributions for associated $H + V$ production
measured by ATLAS and D0 (blue bars), with the LHC TGC data (red lines), and the global combination with
both the associated production and TGC data (black bars)~\protect\cite{ESY4}.}
\label{fig:dim6}
\end{figure}

Since the new particle discovered in 2012 looks so much like the Higgs boson of the Standard Model,
a popular way to parametrize our ignorance what may lie beyond it
is to assume that the Higgs and all the other Standard Model particles have exactly the
Standard Model interactions, supplemented by higher-dimensional operators $O_i$ composed of these
Standard Model fields:
\beq
\Delta {\cal L} \; = \; \sum_i C_i O_i \, ,
\label{dim6}
\eeq
where the unknown coefficients $C_i$ have mass dimensions 4 - [Dim $O_i$] and hence are likely
to be suppressed by the corresponding powers of some higher mass scale. Expressing these
coefficients as ${\bar c}_i/m_W^2$, one can use present data (precision electroweak data, Higgs
data, triple-gauge couplings (TGCs), etc.) to constrain the dimensionless reduced coefficients $c_i$.
The results of one such analysis of dimension-6 operators are shown in Fig.~\ref{fig:dim6}: there are no indications that
any deviate significantly from zero. The Standard Model still rules OK!

\section{Apr\`es Higgs}

The discovery of the Higgs boson marks a watershed in particle physics.
In the future, the calendar of particle physics will surely be divided into BH (before Higgs) and AH (after Higgs), with
2012 being year 0. The Higgs boson will signpost the direction that both theoretical and experimental
physics will take in the decades to come.

\begin{figure}[htb]
\centering
\includegraphics[height=2in]{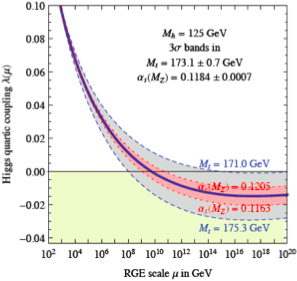}
\hspace{0.5cm}
\includegraphics[height=2in]{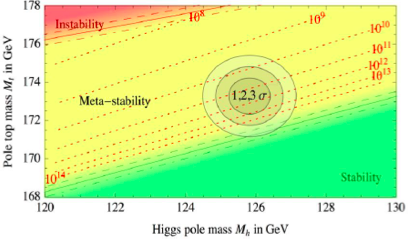}
\caption{\it Left panel: Within the SM, renormalization by the top quark appears to drive the Higgs self-coupling $\lambda < 0$ at
large scales, destabilizing the electroweak vacuum. Right panel: Regions of vacuum stability, metastability and instability in the $(m_H, m_t)$ plane.
Taken from~\protect\cite{Buttazzo}.}
\label{fig:Buttazzo}
\end{figure}

We can be optimistic, because there are already many arguments that
there must be new physics. One of them is provided by measured value of the Higgs mass (\ref{measuredmH}).
Coupled with the world average value of the mass of the top quark: $m_t = 173.34 \pm 0.76$~GeV.
Extrapolating to high renormalization scales indicates~\cite{Buttazzo} that the effective Higgs quartic coupling $\lambda$
would become negative at a Higgs scale $\Lambda$, as seen in the left panel of Fig.~\ref{fig:Buttazzo}:
\begin{eqnarray}
\log_{10} \left( \frac{\Lambda}{{\rm GeV}} \right) & = & 11.3 + 1.0 \left(\frac{m_H}{{\rm GeV}} - 125.66 \right) \nonumber \\
& - & 1.2 \left( \frac{m_t}{{\rm GeV}} - 173.10 \right) \nonumber \\
& + & 0.4 \left(\frac{\alpha_s(M_Z) - 0.1184}{0.0007} \right) \, .
\label{Buttazzo}
\end{eqnarray}
Using the world average values of $m_t$, $m_H$ and $\alpha_s (M_Z) = 0.1185 \pm 0.0006$, this formula yields
\begin{equation}
\Lambda \; = \; 10^{10.6 \pm 1.0}~{\rm GeV} \, .
\label{Lambda}
\end{equation}
As seen in the right panel of Fig.~\ref{fig:Buttazzo}, the most important uncertainty in this calculation is that associated with $m_t$, but all the indications
are that there is an instability in the electroweak vacuum. The lifetime for tunnelling from our present
electroweak vacuum to a state with a Higgs v.e.v. larger than (\ref{Lambda}) is probably longer than the
age of the Universe, so one might be tempted to ignore this problem.
However, if the Universe was once very hot and dense as suggested by conventional Big Bang theory
and cosmological inflation, most of the early Universe would have got stuck in this unphysical state, and
would never have reached the present electroweak vacuum. This indicates that there should be some
new physics below the scale (\ref{Lambda}) that is capable of stabilizing the electroweak vacuum,
such as supersymmetry~\cite{ER}. This is a third reason that Run~1 of the LHC has given us to favour supersymmetry,
in addition to its successful prediction of the Higgs mass and the similarity of its couplings to those in the Standard Model.

There are many other reasons why there must be physics beyond the Standard Model, and the Higgs boson
may play an important r\^ole in most of them. What is the origin of the matter in the Universe - is it due to
a first-order electroweak phase transition? What is the dark matter - could the Higgs boson serve as a portal
towards it? How to stabilize the Higgs mass and hence the electroweak scale so far below the Planck scale?
Are neutrino masses due to the conventional Higgs, or some different mechanism? What is the dark energy, and why
is it much smaller than the electroweak scale? Is the great size and age of the Universe, and its near-flatness, due
to a primordial epoch of cosmological inflation, driven by the energy in some scalar field like the Higgs, perhaps
even the Higgs itself?

With the discovery of the Higgs boson in 2012, many new directions for physics have
opened up. On the one hand, there is a need for detailed investigation of the Higgs, to see whether
it conforms to the Standard Model paradigm or whether it exhibits deviations due to new physics.
On the other hand, the hunt will be on for whatever new physics complements the Higgs boson, be it
supersymmetry or ...? We look forward to the
years AH {\it (Anno Higgsi)} $> 0$.


\section*{Acknowledgements}

The work of J.E. was supported in part by
the London Centre for Terauniverse Studies (LCTS), using funding from
the European Research Council 
via the Advanced Investigator Grant 267352 and from the STFC
(UK) via the research grants ST/J002798/1 and ST/L000326/1. 
The work of M.K.G. was supported in part by the Director, Office of Science,
Office of High Energy and Nuclear Physics, Division of High Energy
Physics, of the U.S. Department of Energy under Contract
DE-AC02-05CH11231, in part by the National Science Foundation under
grant PHY-1002399.
The work of D.V.N. was supported in part by the
DOE grant DE-FG02-13ER42020.


\end{document}